\documentstyle[12pt,dina4p]{article}
\newcommand{\nc}{\newcommand}
\nc{\rnc}{\renewcommand}
\nc{\bm}{\bibitem}
\def\3{\ss}
\nc{\mb}{\mbox}
\nc{\be}{\begin{equation}}
\nc{\ee}{\end{equation}}
\nc{\bea}{\begin{eqnarray*}}
\nc{\beam}{\begin{eqnarray}}
\nc{\eea}{\end{eqnarray*}}
\nc{\eeam}{\end{eqnarray}}
\nc{\nn}{\nonumber}
\nc{\dps}{\displaystyle}
\nc{\nichts}{\rule{0ex}{1ex}}
\nc{\wenig}{\hspace*{1cm}}
\nc{\etwas}{\hspace*{2cm}}
\nc{\mehr}{\hspace*{3cm}}
\nc{\Rz}{Raum-Zeit}
\nc{\ped}{positive-Energie-Darstellung}
\nc{\hkn}{Haag-Kast\-ler-Netz}
\nc{\nel}{Netze auf dem eindimensionalen Lichtkegel}
\nc{\npf}{$n$-Punkt-Funktion}
\nc{\zpf}{2-Punkt-Funktion}
\nc{\cst}{$C^{\ast}\!$-Al\-ge\-bra}
\nc{\csts}{$C^{\ast}\!$-Al\-ge\-bren}
\nc{\fO}[1]{\mb{${\cal#1(O)}$}}
\nc{\NO}[1]{\mb{${\cal A:=\overline{\bigcup_{O}#1(O)}^{\|\cdot\|}}$}}
\nc{\fOO}[1]{\mb{${\cal #1 : O \rightarrow #1(O)} $}}
\nc{\fOOO}[1]{\mb{${\cal #1}: I \rightarrow{\cal #1}(I)$\,, $I\in\K$}}
\nc{\FOOO}[1]{\mb{${\cal #1}: I \rightarrow{\cal #1}(I)$\,, $I\in\KK$}}
\nc{\A}{\mb{${\cal A}$}}
\nc{\B}{\mb{${\cal B}$}}
\nc{\BH}{\mb{$B(H)$}}
\nc{\C}{\mb{\bf C}}
\nc{\CC}{\mb{${\cal C}$}}
\nc{\D}{\mb{${\cal D}$}}
\rnc{\d}{\mb{d}}
\nc{\dd}{\!\mb{{\footnotesize {\it d}}}}
\nc{\e}{\mb{$\varepsilon$}}
\nc{\F}{\mb{${\bf{\cal F}}$}}
\nc{\fn}{\mb{$F_n$}}
\nc{\Fn}{\mb{$\overline{F}_n$}}
\nc{\G}{{\it G}}
\rnc{\H}{{\it H}}
\nc{\HH}{\mb{{\bf H}}}
\rnc{\k}{\mb{${\cal K}_0$}}
\nc{\K}{\mb{${\cal K}$}}
\nc{\KK}{\mb{$\overline{\K}$}}
\nc{\KKK}{\mb{\scriptsize ${\cal K}_0$}}
\rnc{\l}{\mb{$\lambda$}}
\nc{\N}{\mb{${\bf N}$}}
\rnc{\o}{\mb{$\omega$}}
\nc{\OM}{\mb{$\Omega$}}
\rnc{\O}{\mb{${\cal O}$}}
\rnc{\P}{\mb{{\bf P}}}
\nc{\R}{\mb{\bf R}}
\nc{\Rp}{\mb{$\R_{+}$}}
\nc{\Rm}{\mb{$\R_{-}$}}
\rnc{\r}{\mb{\scriptsize \bf R}}
\nc{\rp}{\mb{$\r_{\mb{\tiny +}}$}}
\nc{\rmm}{\mb{$\r_-$}}
\rnc{\S}{\mb{${\cal S}$}}
\nc{\SL}{\mb{$SL(2,\R)/\Z_{2}$}}
\nc{\SO}{\mb{${\cal S(O)}$}}
\nc{\SpP}{\mb{Sp\,{\bf P}}}
\nc{\SR}{\mb{${\cal S(\R)}$}}
\rnc{\t}{\mb{$\vartheta$}}
\nc{\U}{{\it U}}
\rnc{\v}{\mb{$\varphi$}}
\nc{\vg}{\overline{\v_g}}
\nc{\Z}{\mb{\bf Z}}
\title{From Conformal Haag-Kastler Nets\\ to Wightman Functions\footnote{hep-th/9609020}}
\author{Martin J\"or\ss\\
II. Institut f\"ur theoretische Physik der Universit\"at Hamburg\\
Luruper Chaussee 149, 22761 Hamburg\\
e-mail: joerss@x4u2.desy.de}
\date{August 1996}
\begin{document}
\maketitle
\begin{abstract}
Starting from a chiral conformal Haag-Kastler net on 2 dimensional 
Minkowski space we present a canonical construction that leads to 
a complete set of conformally covariant
 N-point-functions fulfilling the Wightman 
axioms. 

Our method consists of an explicit use of the representation theory of
the universal covering group of $SL(2,{\bf R})$ combined with 
a generalization of the conformal cluster theorem to N-point-functions \cite{FrJ}.

This paper continues work done in \cite{FrJ} and \cite{Joe3}.
\end{abstract}
\section{Introduction}
The formulation of quantum field theory in terms of
Haag Kastler nets of local observable algebras
(``local quantum physics" \cite{Haa}) has
turned out to be well suited for
the investigation of general structures.
Discussion of concrete models, however, is mostly done
in terms of pointlike localized fields.

In order to be in a precise mathematical framework,
these fields might be assumed to obey the Wightman axioms \cite{StW}.
Even then, the interrelation between both concepts is not
yet completely understood (see \cite{BaW,BoY} for the present stage).

Heuristically, Wightman fields are constructed out
of Haag-Kastler nets by some scaling limit which, however, is
difficult to formulate in an intrinsic way \cite{Buc2}.
                                           In a dilation invariant
theory scaling is well defined,
and in the presence of massless particles the construction
of a pointlike field was performed in \cite{BuF}.

Here, we study the possibly simplest
situation: Haag-Kastler nets in 2 dimensional Min\-kows\-ki
space with trivial translations in one light cone
direction (``chirality") and covariant under
the real M\"obius group which acts on the other lightlike
direction. 

In \cite{FrJ}, it has been shown that in the vacuum representation
pointlike localized fields can be constructed. Their
smeared linear combinations are affiliated to the original
net and generate it. We do not know at the moment whether
they satisfy all Wightman axioms,
since we have not yet found
      an invariant domain of definition.

In \cite{Joe3}, we have generalized this to the charged sectors of a
theory. We have constructed pointlike localized fields carrying
arbitrary charge with finite statistics
                 and therefore intertwining between the different
superselection sectors of the theory.
(In Conformal Field Theory
 these objects are known as ``Vertex Operators".)
                                      We have obtained
                                          the unbounded field
operators as limits of elements of the reduced field bundle
                             \cite{FRS1,FRS2}
                                                            associated
to the net of observables of the theory.

In this paper, we start again from chiral conformal Haag-Kastler nets
and present an canonical construction of
N-point-functions that can be shown to fulfill the Wightman axioms. 
We proceed by generalizing the conformal cluster theorem \cite{FrJ} to
higher N-point-functions and by examining the momentum space limit of
the algebraic N-point-functions at $p=0$.

We are not able to prove that these Wightman fields can be identified
with the pointlike localized fields constructed in \cite{FrJ} and \cite{Joe3}.
\section{First Steps}
In this section, we give an explicit formulation
of the setting frow which this work starts. We then present the proof
of the conformal cluster theorem and the results on the construction
of pointlike localized fields in \cite{FrJ} and \cite{Joe3}.
\subsection{Assumptions}
Let $\A=(\A(I))_{I\in\KKK}$
                       be a family of von Neumann algebras on some
separable Hilbert space \H. \k\ denotes the set of nonempty bounded
open intervals on \R.
$\A$ is assumed to satisfy the following conditions.
\begin{enumerate}
\def\labelenumi{\roman{enumi})}
\def\theenumi{\roman{enumi}}
\item Isotony:
\be \A(I_1)\subset\A(I_2)\;\;\;\;\;\mb{for}\;\;\;\;I_1\subset I_2,
\;\;\;\;I_1, I_2\in\k.\ee
\item Locality:
\be \A(I_1)\subset\A(I_2)'\;\;\;\;\;\mb{for}\;\;\;\;I_1\cap I_2=\{\},
\;\;\;\;I_1, I_2\in\k\ee
($\A(I_2)'$ is the commutant of $\A(I_2)$).
\item
There exists a strongly continuous unitary representation $U$ of
$G=SL(2,\R)$ in $H$ with $U(-1)=1$ and
\be U(g)\,\A(I)\,U(g)^{-1}=\A(gI),\;\;\;\;\;I,gI\in\k
\ee
($SL(2,\R)\ni g=\left(\begin{array}{cc}a&b\\c&d\end{array}\right)$
                       acts
                            on $\R\cup\{\infty\}$ by
$x\mapsto
\frac{ax+b}{cx+d}$ with the appropriate interpretation for
$x, gx=\infty$).
\item
The conformal Hamiltonian \HH, which generates the
 restriction of $U$ to $SO(2)$, has a nonnegative spectrum.
\item
There is a unique (up to a phase) $U$-invariant unit vector
$\OM\in\H$.
\item
\H\ is the smallest closed subspace containing the vacuum
                                               $\OM$ which
is invariant under $U(g),$ $g\in SL(2,\R),$ and $A\in\A(I),I\in\k$
(``cyclicity").
\footnote{This assumption is seemingly weaker than cyclicity of
$\OM$ w.r.t.\,the algebra of local observables on $\R$.}
\end{enumerate}
It is convenient to extend the net to intervals $I$ on the circle
$S^1=\R\cup\{\infty\}$ by setting
\be
\A(I)=U(g)\;\A(g^{-1}I)\;U(g)^{-1},\;\;\;\;\;\;g^{-1}I\in\k,\;g\in
SL(2,\R).
\ee
The covariance property guarantees that $\A(I)$ is well defined
for all intervals $I$ of the form $I=gI_0,\;I_0\in\k,\;g\in SL(2,\R),$
i.e.\ for all nonempty nondense open intervals on $S^1$ (we denote
the set of these intervals by \K).
\subsection{Conformal Cluster Theorem}
In this subsection, we derive a bound on conformal two-point-functions
in algebraic quantum field theory (see \cite{FrJ}). This bound specifies the
decrease properties of conformal two-point-functions in
the algebraic framework to be exactly those known from theories
with pointlike localization. The Conformal Cluster Theorem plays a central role in this work.

\medskip

{\bf Conformal Cluster Theorem (see \cite{FrJ}):}
Let $(\A(I))_{I\in\KKK}$ be a conformally covariant local net on $\R$.
Let       $a,b,c,d\in\R$ and
              $a<b<c<d$.
Let $A\in\A(\,(a,b)\,)$, $B\in\A(\,(c,d)\,)$, $n\in\N$ and
$P_k A\OM=P_k A^*\OM=0,\;k<n$. $P_k$ here denotes the
projection on the subrepresentation of $U(G)$ with conformal dimension $k$.
We then have
\be
|(\,\OM,BA\OM\,)|\leq \left(\frac{(b-a)\,(d-c)}{(c-a)\,(d-b)}\right)^n
\;\|A\|\,\|B\|.
\ee

\medskip

{\bf Proof:} Choose $R>0$. We consider the following 1-parameter
subgroup of $G=SL(2,\R)$\,:
\be
g_t\,:\,x\longmapsto\frac{x\,
                       \mb{cos}\frac{t}{2}+R\,\mb{sin}\frac{t}{2}}
{-\frac{x}{R}\,\mb{sin}\frac{t}{2}+\mb{cos}\frac{t}{2}}\,.
\ee
Its generator ${\rm \bf H}_R$ is within each subrepresentation
of $U(G)$
                                 unitarily equivalent to the
conformal Hamiltonian ${\rm \bf H}$. Therefore, the spectrum of $A\OM$
and $A^*\OM$ w.r.t.\ ${\rm \bf H}_R$ is bounded below by $n$.
Let $0<t_0<t_1<2\pi$ such that $g_{t_0}(b)=c$ and $g_{t_1}(a)=d$.
We now define
\be
F(z)=\left\{\begin{array}{ll}
                           (\,\OM,\,B\,z^{-{\rm \bf H}_R}\,A\OM\,)
&|z|>1\\
(\,\OM,\,A\,z^{{\rm \bf H}_R}\,B\OM\,)&|z|<1\\
(\,\OM,\,A\,\alpha_{g_t}(B)\,\OM\,)&z=e^{it},\,t\not\in
[t_0,t_1]
\end{array},
\right.
\ee
a function analytic in its domain of definition, and then
\be
G(z)=\,(z-z_0)^n\,(z^{-1}-z_0^{-1})^n\,F(z),\;\;
                         z_0=e^{\frac{i}{2}(t_0+t_1)}\,.
\ee
(Confer the idea in \cite{Fre}.)
At $z=0$ and $z=\infty$ the function
                   $G(\cdot)$ is bounded
because of the bound on the spectrum of ${\rm \bf H}_R$ and
can therefore be analytically continued.
As an analytic function it reaches its maximum at the boundary of its
domain of definition, which is the interval
$[e^{it_0},e^{it_1}]$
on the unit circle:
\be
\mb{sup}|G(z)|\,\leq\,\|A\|\,\|B\|\,|e^{it_0}-e^{\frac{i}{2}
(t_0+t_1)}|^{2n}\,=\,\|A\|\,\|B\|\,(2\,
                                     \mb{sin}\frac{t_0-t_1}{4})^{2n}
\,.
\ee
This leads to
\beam
|(\,\OM,\,BA\OM\,)|&=&|F(1)|\,=\,|G(1)|\ |1-e^{\frac{i}{2}(t_0+t_1)}
                                                              |^{
-2n}\,=\,|G(1)|\ (2\,\mb{sin}\frac{t_0+t_1}{4})^{-2n}\nn\\
&\leq&\mb{sup}|G|\,(2\,\mb{sin}\frac{t_0+t_1}{4})^{-2n}\,\leq\,
\|A\|\,\|B\|\,\left(\frac{\mb{sin}\frac{t_0-t_1}{4}}{\mb{sin}
\frac{t_0+t_1}{4}}\right)^{2n}\,.
\eeam
Determining $t_0$ and $t_1$ we obtain
\be
\lim_{R\rightarrow\infty}R\,t_0=2(c-b)\;\;\;\mb{and}\;\;\;
\lim_{R\rightarrow\infty}R\,t_1=2(d-a)\,.
\ee
We now assume $a-b=c-d$ and find $\left(\frac{t_0-t_1}{t_0+t_1}
\right)^2=\frac{(a-b)\,(c-d)}{(a-c)\,(b-d)}=:x\,.$
Since the bound on $|(\,\OM,\,BA\OM\,)|$ can only depend
on the conformal cross ratio $x$,
we can drop the assumption and the theorem is
proven.\,\hfill $\Box$
\subsection{The Construction of Pointlike Localized
  Fields from Conformal Haag-Kastler Nets}
This subsection presents
the argumentation and results of \cite{FrJ} and \cite{Joe3}:\\
The idea for the definition of conformal fields is the following:
Let $A$ be a local observable,
\be
A\in\bigcup_{I\in\KKK}\A(I),
\ee
and $P_\tau$ the projection onto an irreducible subrepresentation
$\tau$ of $U$. The vector $P_\tau A\OM$ may then be thought of as
$\varphi_\tau(h)\,\OM$ where $\varphi_\tau$ is a conformal field of
dimension $n_\tau=:n$ and $h$ is an appropriate
                                function on $\R$.
                                           The relation between
$A$ and $h$, however, is unknown at the moment, up to the known
transformation properties under $G$,
\be
U(g)\,P_\tau A\OM=\varphi_\tau(h_g^{(n)})\,\OM
\ee
with $h_g^{(n)}(x)=(cx-a)^{2n-2}\,h(\frac{dx-b}{-cx+a})$,
$g=\left(\begin{array}{cc}a&b\\c&d\end{array}\right)\in G$.
We may now scale the vector $P_\tau A\OM$ by dilations $D(\l)=
U\left(\begin{array}{cc}\l^{\frac{1}{2}}&0\\0&\l^{-\frac{1}{2}}
\end{array}\right)$ and find
\be
D(\l)\,P_\tau A\OM=\l^n\,\varphi_\tau(h_{\l})\,\OM
\ee
where $h_{\l}(x)=\l^{-1}\,h(\frac{x}{\lambda
                                     })$. Hence, we obtain formally
for $\lambda\downarrow 0$
\be
\l^{-n}\,D(\l)\,P_\tau A\OM\longrightarrow
                                          \int dx\,h(x)\;\varphi_\tau(0)
\,\OM.
\ee
In order to obtain a Hilbert space vector in the limit, we smear over
the group of
    translations $T(a)=U\left(\begin{array}{cc}1&a\\0&1\end{array}
\right)$ with some test function $f$ and obtain formally
\be
\label{a}
\lim_{\lambda
        \downarrow 0}\l^{-n}\int da\,f(a)\;T(a)\,D(\l)\,P_\tau A\OM=
\int dx\,h(x)\;\varphi_\tau(f)\,\OM.\label{f}
\ee
We now interpret the left-hand side as a definition of a conformal
field $\varphi_\tau$ on the vacuum, and try to obtain densely defined
operators with the correct localization by defining
\be
\varphi_\tau^I(f)\,A'\OM=A'\varphi_\tau^I(f)\,\OM,\;\;
f\in\D(I),\,A'\in\A(I)',\,I\in\K.
\ee
In order to make this formal construction meaningful, there are two problems to overcome.

The first one is the fact that the limit on the left-hand side of
(\ref{a})
     does not exist in general if $A\OM$ is replaced by an arbitrary
vector in $H$. This corresponds to the possibility that the function
$h$ on the right-hand side might not be integrable.
We will show that after smearing the operator $A$
                                          with a smooth function on $G$,
the limit is well defined. Such operators will be called regularized.
The second problem is to show that the smeared field operators
$\varphi_\tau^I(f)$ are closable, in spite of the nonlocal
nature of the projections $P_\tau$. 

We omit the technical parts of \cite{FrJ} and \cite{Joe3} and
summarize the results in a compact form and as general as possible.

Due to the positivity condition, the representation
              $U(\tilde{G})$ is completely
reducible into irreducible subrepresentations
         and the irreducible components $\tau$ are up to equivalence
uniquely characterized by the conformal dimension $n_\tau\in\Rp$
($n_\tau$ is the lower bound of the spectrum of the conformal
 Hamiltonian \HH\ in the representation $\tau$).

Associated with each irreducible subrepresentation $\tau$ of $U$
we find
for each $I\in\k$ a
     densely defined operator-valued distribution $\varphi_\tau^I$
on the space $\D(I)$
             of Schwartz functions with support in $I$
   such that the following statements hold for all $f\in\D(I).$
\begin{enumerate}
\def\labelenumi{\roman{enumi})}
\def\theenumi{\roman{enumi}}
\item
The domain of definition of
$\varphi_\tau^I(f)$ is given by $\A(I')\,\OM$.
\item
\be
\varphi_\tau^I(f)\,\OM\in P_\tau\H_{red}
\ee
with $P_\tau$ denoting the projector on the module of $\tau$.
\item
\be
U(\tilde{g})\;
    \varphi_\tau^I(x)\;
        U(\tilde{g})^{-1}=(cx+d)^{-2n_\tau}\varphi_\tau^{gI}
                                 (\tilde{g}x)
\ee
with the covering projection $\tilde{g}\mapsto g$ and
$g=\left(\begin{array}{cc}a&b\\c&d\end{array}\right)\in
SL(2,\R),\;I, gI\in\k$.
\item
$\varphi_\tau^I(f)$ is closable.
\item
The closure of $\varphi_{\tau}^I(f),\,f\in\D(I),$ is affiliated to
                                          $\A(I)$.
\item
$\A(I)$ is the smallest von Neumann algebra to which all operators
$\varphi_{\tau}^I(f),\,f\in\D(I),$ are affiliated.
\item The exchange algebra of the reduced field bundle \cite{FRS2}
and the existence of the closed field operators
$\varphi_\tau^I(f)$, mapping a dense set of the vacuum Hilbert space
into some charged sector with finite statistics,
                          suffice to construct  closed field operators
$\varphi_{\tau,\alpha}^I(f)$,
                             mapping a dense set of an arbitrary charged
sector $\alpha$ with finite statistics
                into
                     some (other) charged sector with finite statistics.
Here,
                    the irreducible
                                          module $\tau$ of
$U(\tilde{G})$ labels orthogonal irreducible fields defined
in the same sector $\alpha$\,.
\item
The closure of any $\varphi_{\tau,\alpha}^I(f),\,f\in\D(I),$ is affiliated to
                                          $\F_{red}(I)$.
\item
$\F_{red}(I)$ is the smallest von Neumann algebra to which all operators
$\varphi_{\tau,\alpha}^I(f),\,f\in\D(I),$ are affiliated.
\end{enumerate}

With the existence of pointlike localized fields we are able to
give a proof of a generalized Bisognano-Wichmann property. We can
identify the conformal group and the reflections as generalized
modular structures in the reduced field bundle. Especially, we obtain
a PCT operator on $\H_{red}$ proving the PCT theorem for the full
theory.

Moreover, the existence of pointlike localized fields gives a
proof of the hitherto unproven Spin-Statistics theorem for conformal
Haag-Kastler nets in 1+1 dimensions.

It was also possible to prove an operator product expansions
for arbitrary local observables: \\
For each $I\in\k$ and each $A\in \A(I)$
there is a local expansion
\be
A\,=\,\sum_{\tau}\varphi_{\tau}^I(f_{\tau,A})\,
\ee
into a sum over all irreducible modules
$\tau$ of $U(G)$
with \be
 \mb{supp}f_{\tau,A}\subset I\,,\ee
which converges on $\A(I')\OM$ $*$-strongly (cf.\ the definition in
\cite{BrR}). Here, $I'$ denotes the complement of $I$ in $\k$.
\section{Canonical Construction of Wightman Fields}
Starting from a chiral
conformal Haag-Kastler net, pointlike localized fields have been
constructed in \cite{FrJ,Joe3}. Their smeared linear combinations are affiliated to the
original net and generate it. We do not know at the moment whether
these fields satisfy all Wightman axioms, since we have not found an
invariant domain of definition.

In this section, we construct in a canonical manner a complete set of
pointlike localized correlation functions out of
the net of algebras we have been starting from. 
We proceed by generalizing the conformal cluster theorem to
higher N-point-functions and by examining the momentum space limit of
the algebraic N-point-functions at $p=0$. This canonically constructed
set of correlation
functions can be shown to fulfill the conditions for Wightman functions (cf.\ \cite{StW} and
\cite{Jos}). Hence, we can construct an associated
field theory fulfilling the Wightman axioms. 

We are not able to prove that these Wightman fields can be identified with the pointlike
localized fields constructed in \cite{FrJ} and \cite{Joe3}. We do not know
either how the Haag-Kastler theory, we have been starting
from, can be reconstructed from the Wightman theory. 

Such phenomena
have been investigated by Borchers and Yngvason \cite{BoY}. Starting
from a Wightman theory, they could not rule out in general the possibility that
the associated local net has to be defined in an enlarged Hilbert space.
\subsection{Conformal Two-Point-Functions}
First, we will determine the general form of conformal
two-point-functions of local observables:\\
It has been shown (cf.\ e.g.\ \cite{Joe1}) that a two-point-function
$(\,\OM,\,B\,U(x)\,A\OM\,)$
of a chiral local net with translation covariance  is of
Lebesgue class $L^p$ for any $p>1$. The
           Fourier
transform of this two-point-function is a measure concentrated on
the positive half line. Therefore, it is - with the possible
                             exception of a trivial
delta function at zero - fully determined by the Fourier transform
of the commutator function $(\,\OM,\,
[B,\,U(x)\,A\,U(x)^{-1}]\,\OM\,)\,.$ Since $A$ and $B$ are local
observables, the commutator function has compact support and an
analytic Fourier transform $G(p)$.
                           The restriction $\Theta(p)\,G(p)$
                                           of this analytic
function to the positive half line is then the Fourier transform
of $(\,\OM,\,B\,U(x)\,A\OM\,)\,.$

In the conformally covariant case with
$P_kA\OM=P_kA^*\OM=0,\,k<n$, the
conformal cluster theorem implies that the
two-point-function
$(\,\OM,\,B\,U(x)\,A\OM\,)$
decreases as $x^{-2n}$. Therefore,
its Fourier transform is $2n\!-\!2\,$times continuously differentiable
and can be written as $\Theta(p)\,p^{2n-1}\,H(p)$ with an
appropriate analytic function $H(p)$.

Using this result, we are able to present a sequence of canonically
scaled two-point-functions of local observables
converging as distributions to the two-point-function known from
conventional conformal field theory (cf.\ \cite{Joe1,Reh}):
\be 
\lim_{\lambda\downarrow 0}\l^{-2n}\ (\,\OM,\,B\,U(
\lambda^{-1}x)\,A\OM\,)\ =\ 
\lim_{\lambda\downarrow 0}\l^{-2n}\ {\cal F}_{p\rightarrow x}\
 \Theta(p)\,(\l p)^{2n-1}\,H(\l p)\,\l\,dp\ =\ 
H(0)\ (x+i\varepsilon)^{-2n}\,.
\ee
\subsection{Conformal Three-Point Functions}
We consider the properties of chiral algebraic three-point functions 
\be
(\,\OM,\,A_1\,U(x_1-x_2)\,A_2\,U(x_2-x_3)\,A_3\,\OM\,)
\ee
 of local observables
$A_i$\,, $ i=1,2,3$\,. 
 The general form of a
(truncated)  chiral three-point function of local observables
is restricted by locality and by the condition of positive energy. The
Fourier transform of an algebraic three-point function
 can be shown to be the  sum of the restrictions of analytic functions to
 disjoint open wedges in the domain of positive energy:
\\  If $F$ now denotes the Fourier
 transform of $(\,\OM,\,A_1\,U(\cdot)\,A_2\,U(\cdot)\,A_3\,\OM\,)$,
  we get by straightforward calculations as a first result (cf.\ \cite{Joe4})
\be
F(p,q)\,=\,\Theta(p)\,\Theta(q-p)\,G^+(p,q)\,+\,
\Theta(q)\,\Theta(p-q)\,G^-(p,q)
\ee
with appropriate analytic functions
$G^+$ and $G^-$.
 
In the case of conformal covariance the general
form of these algebraic three-point functions is even more restricted by the
following generalization of the conformal cluster theorem \cite{FrJ}:

\medskip

{\bf Theorem:} Let $(\A(I))_{I\in\KKK}$ be a conformally covariant local net on $\R$\,.
Let       $a_i,b_i\in\R\,,\ i=1,2,3$\,, and
              $a_1<b_1<a_2<b_2<a_3<b_3$\,.
Let $A_i\in\A(\,(a_i,b_i)\,)$\,, $n_i\in\N$\,, $i=1,2,3$\,, and 
\be
P_k\,A_i\,\OM=P_k\,A_i^*\,\OM=0\,,\;k<n_i\,.
\ee
 $P_k$ here denotes the
projection on the subrepresentation of $U(SL(2,R))$ with conformal dimension $k$\,.
We then have the following bound:
\beam
|(\,\OM,A_1A_2A_3\OM\,)|&\leq& 
\left|\frac{(a_1-b_1)+(a_2-b_2)}{(a_2-a_1)+(b_2-b_1)}\right|^{(n_1+n_2-n_3)}\\
&&
\left|\frac{(a_1-b_1)+(a_3-b_3)}{(a_3-a_1)+(b_3-b_1)}\right|^{(n_1+n_3-n_2)}
\;\;\nn\\&&
\left|\frac{(a_2-b_2)+(a_3-b_3)}{(a_3-a_2)+(b_3-b_2)}\right|^{(n_2+n_3-n_1)}
\;\|A_1\|\,\|A_2\|\,\|A_3\|\,.\nn
\eeam
If we additionally assume 
\be
a_1-b_1=a_2-b_2=a_3-b_3\,,
\ee
 we get 
\be
|(\,\OM,A_1A_2A_3\OM\,)|\ \leq\
r_{12}^{(n_1+n_2-n_3)/2}\,r_{23}^{(n_2+n_3-n_1)/2}\,
r_{13}^{(n_1+n_3-n_2)/2}\ \|A_1\|\,\|A_2\|\,\|A_3\|\,,
\ee
with the conformal
cross ratios
\be
\frac{(a_i-b_i)\,(a_j-b_j)}{(a_i-a_j)\,(b_i-b_j)}=:r_{ij}\,,\,i,j=1,2,3\,.
\ee

\medskip

{\bf Proof:}  This proof follows, wherever possible, the line of
argument in the proof of the conformal cluster theorem for 
two-point functions (cf.\ \cite{FrJ}). \\
Choose $R>0$\,. Let us consider the following one-parameter
subgroup of $SL(2,\R)$\,:
\be
g_t\,:\,x\longmapsto\frac{x\,
                       \mb{cos}\frac{t}{2}+R\,\mb{sin}\frac{t}{2}}
{-\frac{x}{R}\,\mb{sin}\frac{t}{2}+\mb{cos}\frac{t}{2}}\,.
\ee
Its generator ${\rm \bf H}_R$ is within each subrepresentation
of $U(SL(2,R))$
                                 unitarily equivalent to the
conformal Hamiltonian ${\rm \bf H}$\,. Therefore, the spectrum of $A_i\,\OM$
and $A^*_i\,\OM$ with respect to ${\rm \bf H}_R$ is bounded from below by $n_i$\,, $i=1,2,3$\,.
Let $0<t^-_{ij}<t^+_{ij}<2\pi$ such that
 \be
g_{t^-_{ij}}(b_i)=a_j
\ee
 and 
\be
g_{t^+_{ij}}(a_i)=b_j\,
\ee
for $i,j=1,2,3$\,, $i<j$\,.
We now define
\be
F(z_1,z_2,z_3):=(\,\OM,\,A_{i_1}\,(\frac{z_{i_1}}{z_{i_2}})^{{\rm \bf H}_R}\,A_{i_2}\,(\frac{z_{i_2}}{z_{i_3}})^{{\rm \bf H}_R}\,A_{i_3}\OM\,)
\ee
in a domain of definition given by
\be
|z_{i_1}|<|z_{i_2}|<|z_{i_3}|
\ee
 with permutations $(i_1,i_2,i_3)$
of $(1,2,3)$\,. This definition can uniquely be extended 
to certain boundary values with $|z_{j}|=|z_{k}|$\,, $j,k=1,2,3$\,, $j\neq
k$\,:\\
$F$ shall be continued to this boundary of its domain of definition 
if 
\be
t_{jk}:=-i\log\frac{z_{j}}{z_{k}}\notin[t^-_{jk},t^+_{jk}]+2\pi{\bf
  Z}  
\ee
or equivalently if
\be
g_{t_k}([a_k,b_k])\cap g_{t_j}([a_j,b_j])\neq\emptyset\,,
\ee
using the notation 
\be
t_i:=-i\log z_i\,,\ i=1,2,3\,.
\ee
Thereby, boundary points with coinciding absolute values are included in the
domain of definition. The definition of $F$ is chosen in
analogy to the analytic continuation of general Wightman functions
(cf., e.g.,
\cite{StW,Jos})
 such that the edge-of-the-wedge theorem for distributions with
 several variables \cite{StW} proves $F$ to be an analytic function:\\
Permuting the local observables $A_i$\,, $i=1,2,3$\,, we have six
three-point functions
\be
(\,\OM,\,A_{i_1}\,U(x_{i_1}-x_{i_2})\,A_{i_2}\,U(x_{i_2}-x_{i_3})\,A_{i_3}\OM\,)\,.
\ee
These six functions have by locality identical values on a domain
\be
E:=\{(y_1,y_2)\in {\bf
  R}^2\,|\,|y_1|>c_1,\,|y_2|>c_2,\,|y_1+y_2|>c_3\}
\ee
 with appropriate $c_1,c_2,c_3\in\Rp$\,. Each single function can be continued analytically by the condition of positive energy
to one of the six disjoint subsets in 
\be
U:={\bf R}^2+iV:=\{(z_1,z_2)\in
{\bf C}^2\,
|\,\mb{Im}z_1\neq 0 \neq
\mb{Im}z_2\,,\,\mb{Im}z_1
+\mb{Im}z_2\neq 0\}\,.
\ee
In this geometrical situation, the edge-of-the-wedge theorem (cf.\
\cite{StW}, theorem 2.14) proves the assumed analyticity of $F$.\\
With the abbreviation
\be
z_{ij}^0:=e^{i(t^-_{ij}+t^+_{ij})/2}\,,\ i,j=1,2,3\,,
\ee 
we then define
\beam
\lefteqn{
G(z_1,z_2,z_3)}\\
&:=&F(z_1,z_2,z_3)\,\prod_{(i,j,k)\in T(1,2,3)}\,(\frac{z_i}{z_j}-z^0_{ij})^{(n_i+n_j-n_k)/2}\,(\frac{z_j}{z_i}-z^0_{ji})^{(n_i+n_j-n_k)/2}\,,\nn
\eeam
where $T(1,2,3)$ denotes the set $\{(1,2,3),\,(1,3,2),\,(2,3,1)\}$\,. The added polynomial in
$z_i$\,, $i=1,2,3$\,, is constructed such that the degree of
the leading terms are restricted by the assumption on the conformal
dimensions of the
three-point function $F$. Also,
using the binomial formula, it can be controlled by straightforward
calculations that no half odd integer exponents
appear after multiplication of the product. Hence,
at $z_i=0$ and $z_i=\infty$\,, $i=1,2,3$\,, the function
                   $G$ is bounded 
because of the bound on the spectrum of ${\rm \bf H}_R$ and
can therefore be analytically continued. We can find estimates on $G$ by the maximum
principle for analytic functions. In order to get the estimate needed in
this proof,
we do not use the maximum principle for several complex variables
\cite{BoM}. Instead, we present an iteration
of the maximum principle argument used in
the proof of the conformal cluster theorem \cite{FrJ} for the single
variables $z_i$\,, $i=1,2,3$\,, of $G(\cdot,\cdot,\cdot)$ and derive a bound on $G(1,1,1)$\,:\\
Applying the line of argument known from the case of the two-point functions
now to $G(\cdot,1,1)$\,, we get the estimate
\beam
|G(1,1,1)|&\leq& \mbox{sup}_{z_1}\,|G(z_1,1,1)|\nn\\
&=& \mbox{sup}_{z_1\in
  B_{\cdot,1,1}}\,|G(z_1,1,1)|\,.
\eeam
The boundary of the domain of definition of the maximal analytical
continuation of $G(\cdot,1,1)$ is here denoted by 
\be
B_{\cdot,1,1}:=\{e^{it}\,|\,t\notin [t_{12}^-,t_{12}^+]\cup
  [t_{13}^-,t_{13}^+]+2\pi{\bf Z}\}\,.
\ee
Applying this argument to $G(z_1,\cdot,1)$, we analogously get the estimate
\beam
|G(z_1,1,1)|&\leq& \mbox{sup}_{z_2}\,|G(z_1,z_2,1)|\nn\\
&=& \mbox{sup}_{z_2\in
  B_{z_1,\cdot,1}}\,|G(z_1,z_2,1)|\,
\eeam
with $B_{z_1,\cdot,1}$ denoting the boundary of the domain of definition of the maximal analytical
continuation of $G(z_1,\cdot,1)$\,. 
Applying this argument finally to $G(z_1,z_2,\cdot)$\,, we analogously get the estimate
\beam
|G(z_1,z_2,1)|&\leq& \mbox{sup}_{z_3}\,|G(z_1,z_2,z_3)|\nn\\
&=& \mbox{sup}_{z_3\in
  B_{z_1,z_2,\cdot}}\,|G(z_1,z_2,z_3)|\,
\eeam
with $B_{z_1,z_2,\cdot}$ denoting the boundary of the domain of definition of the maximal analytical
continuation of $G(z_1,z_2,\cdot)$\,. 
Having iterated this maximum principle argument for the single
variables $z_i$\,, $i=1,2,3$\,, we can combine the derived estimates and get
\be
|G(1,1,1)|\ \leq\  \mbox{sup}_{t_{jk}=-i\log\frac{z_j}{z_k}\notin[t^-_{jk},t^+_{jk}]+2\pi{\bf
  Z}\,,\ j\neq k}\,|G(z_1,z_2,z_3)|\,.
\ee
Hence, the boundary values of
$G$ have to be evaluated on the domain described by
\be
g_{t_k}([a_k,b_k])\cap g_{t_j}([a_j,b_j])\neq\emptyset\,
\ee
with 
$
t_i=-i\log z_i\,,\ i=1,2,3\,.
$
We find the supremum with the same
calculation as in the proof of the conformal cluster theorem above
(cf.\ \cite{FrJ}):
\beam
|G(1,1,1)|\,&\leq&\,\|A_1\|\,\|A_2\|\,\|A_3\|\,\prod_{(i,j,k)\in T(1,2,3)}\,|e^{it^-_{ij}}-e^{i(t^-_{ij}+t^+_{ij})/2}|^{n_i+n_j-n_k}\nn\\&=&\,\|A_1\|\,\|A_2\|\,\|A_3\|\,\prod_{(i,j,k)\in T(1,2,3)}\,|2\,
                                     \mb{sin}\frac{t^-_{ij}-t^+_{ij}}{4}|^{n_i+n_j-n_k}
\eeam
This leads to another estimate:
\beam
|(\,\OM,A_1A_2A_3\OM\,)|&=&\,|F(1,1,1)|\nn\\
&=&\,|G(1,1,1)|\,\prod_{(i,j,k)\in T(1,2,3)}\,|1-e^{i(t^-_{ij}+t^+_{ij})/2}|^{n_i+n_j-n_k}\nn\\
&=&\,|G(1,1,1)|\,\prod_{(i,j,k)\in T(1,2,3)}\,|2\,\mb{sin}\frac{t^-_{ij}+t^+_{ij}}{4}|^{n_i+n_j-n_k}\nn\\
&\leq&\,\|A_1\|\,\|A_2\|\,\|A_3\|\,\prod_{(i,j,k)\in T(1,2,3)}\,\left|\frac{\mb{sin}\frac{t^-_{ij}-t^+_{ij}}{4}}{\mb{sin}
\frac{t^-_{ij}+t^+_{ij}}{4}}\right|^{n_i+n_j-n_k}
\eeam
Determining $t^-_{ij}$ and $t^+_{ij}$\,, we obtain for $i,j=1,2,3$
\be
\lim_{R\rightarrow\infty}R\,t^-_{ij}=2(a_j-b_i)
\ee
and
\be
\lim_{R\rightarrow\infty}R\,t^+_{ij}=2(b_j-a_i)
\ee
and the first bound in the theorem is proven. If we now assume
\be
a_1-b_1=a_2-b_2=a_3-b_3\,,
\ee
 we find 
\be
\left(\frac{t^-_{ij}-t^+_{ij}}{t^-_{ij}+t^+_{ij}}
\right)^2=\frac{(a_i-b_i)\,(a_j-b_j)}{(a_i-a_j)\,(b_i-b_j)}=r_{ij}\,,\;\;\;\,i,j=1,2,3\,,
\ee
and the theorem is
proven.\hfill $\Box$

\medskip

This theorem can be used to get deeper
insight in the form of the Fourier transforms of algebraic 
three-point functions.
As in the case of the two-point functions, we proceed by transferring the decrease properties of the function in
position space into regularity properties of the
Fourier transform in momentum space.
 
In conventional
conformal field theory, the three-point function with
conformal dimensions $n_i\,,\ i=1,2,3$\,, is known up to multiplicities as 
\beam
f_{n_1n_2n_3}(x_1,x_2,x_3)
&=&(x_1-x_2+i\varepsilon)^{-(n_1+n_2-n_3)}\nn\\
&&(x_2-x_3+i\varepsilon)^{-(n_2+n_3-n_1)}\nn\\
&&(x_1-x_3+i\varepsilon)^{-(n_1+n_3-n_2)}
\eeam
(cf.\ \cite{ChH,Reh}).

Its Fourier transform 
\be
\tilde{f}_{n_1n_2n_3}(p,q)\,=:\,\Theta(p)\,\Theta(q)\,Q_{n_1n_2n_3}(p,q)
\ee
can be calculated to be a sum of the restrictions of homogeneous polynomials
$Q^+_{n_1n_2n_3}$ and $Q^-_{n_1n_2n_3}$ of degree
 $n_1+n_2+n_3-2$
to disjoint open wedges $W_+$ and $W_-$ in the domain of
positive energy (cf.\ \cite{Reh}). 
   
By the bound in the cluster theorem above, we know that
a conformally covariant algebraic three-point function 
$(\,\OM,\,A_1\,U(x_1-x_2)\,A_2\,U(x_2-x_3)\,A_3\OM\,)$
of local observables $A_i$ with minimal conformal dimensions
$n_i$\,, 
$i=1,2,3\,,$ 
decreases in position space at
least as fast
as the associated pointlike three-point function $f_{n_1n_2n_3}(x_1,x_2,x_3)$ known from conventional conformal field theory. Hence,
the Fourier transform 
$F_{A_1A_2A_3}(p,q)$ 
of this algebraic
three-point function has to be at least as regular in momentum space
as the Fourier transform $\tilde{f}_{n_1n_2n_3}(p,q)$ of the associated pointlike
three-point function
known from
conventional conformal field theory: \\
Technically, we use a well-known formula
from the theory of Fourier transforms,
\be
{\cal F}(\mbox{Pol}(X)S)=\mbox{Pol}(\frac{\partial}{\partial Y}){\cal F}S\,,
\ee
for arbitrary temperate distributions $S$ and polynomials Pol$(\cdot)$
with a (multi-dimensional)
variable $X$ in position space and an appropriate associated differential operator
$\frac{\partial}{\partial Y}$ in momentum space.
 $\F$ denotes the Fourier transformation from position space to
 momentum space. 

Let now $S$ be the conformally covariant algebraic three-point function
of local observables $A_i$ with minimal conformal dimensions
$n_i$\,, 
$i=1,2,3\,$:
\be 
S\,:=\,(\,\OM,\,A_1\,U(x_1-x_2)\,A_2\,U(x_2-x_3)\,A_3\,\OM\,)\,
\ee
and $X$ be a pair of two difference variables out of $x_i-x_j$\,,
$i,j=1,2,3\,.$ By the
cluster theorem proved above, we can now choose
an appropriate homogeneous polynomial $\mbox{Pol}(X)$ of degree
$n_1+n_2+n_3-4$ such that the product $\mbox{Pol}(X)\,S$ is still
absolutely integrable in position space. 
Using the formula given above, we see that 
$\mbox{Pol}(\frac{\partial}{\partial Y}){\cal F}S$ is continuous and bounded in
momentum space. 
Furthermore, we have already derived the form of the Fourier transform $F$
of an arbitrary (truncated) algebraic three-point function in a chiral
theory to be
\be
F(p,q)\,=\,\Theta(p)\,\Theta(q-p)\,G^+(p,q)\,+\,
\Theta(q)\,\Theta(p-q)\,G^-(p,q)
\ee
with appropriate analytic functions
$G^+$ and $G^-$.
Thereby, we see that 
in the case of conformal covariance with minimal conformal dimensions
$n_i$\,, $i=1,2,3$\,,
the analytic function $G^+$ ($G^-$)
 can be expressed as the product of an appropriate
 homogeneous polynomial $P^+$ ($P^-$) of degree $n_1+n_2+n_3-2$ restricted to the wedge $W_+$
 ($W_-$) and an appropriate
analytic function $H^+$ ($H^-$)\,. 
Hence, we have proved that the Fourier transform $F_{A_1A_2A_3}$ of the algebraic
three-point function
$(\,\OM,\,A_1\,U(x_1-x_2)\,A_2\,U(x_2-x_3)\,A_3\,\OM\,)$ 
can be written as 
\be
F_{A_1A_2A_3}(p,q)\ =\ \Theta(p)\,\Theta(q)\,P_{A_1A_2A_3}(p,q)\,H_{A_1A_2A_3}(p,q)
\ee
with an appropriate homogeneous function $P_{A_1A_2A_3}(p,q)$ of degree $n_1+n_2+n_3-2$ and
an appropriate continuous and bounded function $H_{A_1A_2A_3}(p,q)$\,.

These results suffice to control the pointlike limit of the
considered correlation functions.
Scaling an algebraic
three-point function in a canonical manner, we construct a sequence
of distributions that converges to the three-point function of
conventional conformal field theory:
\beam
&&\nn\\
\lefteqn{ \lim_{\lambda\downarrow 0}\l^{-(n_1+n_2+n_3)}\ (\,\OM,\,A_1\,U(
\frac{x_1-x_2}{\lambda})\,A_2\,U(
\frac{x_2-x_3}{\lambda})\,A_3\,\OM\,)}\nn\\
&&\nn\\
&=&\lim_{\lambda\downarrow 0}\l^{-(n_1+n_2+n_3)}
\ {\cal F}_{p\rightarrow x_1-x_2\atop q\rightarrow x_2-x_3}\
F_{A_1A_2A_3}(\l p,\l q)\,\l^2\,dp\,dq\nn\\
&&\nn\\
&=&\lim_{\lambda\downarrow 0}\l^{-(n_1+n_2+n_3)}\nn\\
&&
\ {\cal F}_{p\rightarrow x_1-x_2\atop q\rightarrow x_2-x_3}\ \Theta(p)\,\Theta(q)\,
\l^{n_1+n_2+n_3-2}\,P_{A_1A_2A_3}(p,q)\,H_{A_1A_2A_3}(\l p,\l
q)\,\l^2\,dp\,dq\nn\\
&&\nn\\
&=&(x_1-x_2+i\varepsilon)^{-(n_1+n_2-n_3)}\nn\\
&&(x_2-x_3+i\varepsilon)^{-(n_2+n_3-n_1)}\nn\\
&&(x_1-x_3+i\varepsilon)^{-(n_1+n_3-n_2)}\ H_{A_1A_2A_3}(0,0)\,.\\
&&\nn
\eeam 
\subsection{Conformal N-Point Functions}
Since the notational expenditure increases strongly as we come to the
construction of higher N-point functions, we concentrate on
qualitatively new aspects not occurring in the case of
two-point functions and three-point functions. These qualitatively new
aspects in the construction of higher N-point functions are related to
the fact that in conventional field theory the form of higher
N-point functions is not fully determined by conformal covariance.
In conventional conformal field theory conformal covariance restricts
the form of correlation functions of field operators
$\varphi_i(x_i)\,,\ i=1,2,...\,,N\,,$ with
conformal dimension $n_i$ in the following manner (cf.\ \cite{ChH,Reh}):  
\be
 (\OM,\,\left(\prod_{1\leq i\leq N}\,\varphi_i(x_i)\right)\,\OM)\
=\ \left(\prod_{1\leq i<j\leq N}\frac{1}{(x_j-x_i+i\e)^{c_{ij}}}\right)\;f(r_{t_1u_1}^{v_1s_1},...\,,r_{t_{N-3}u_{N-3}}^{v_{N-3}s_{N-3}})\,.\ee
Here,  $f(\cdot,...\,,\cdot)$
denotes an appropriate function
depending on $N\!-\!3$ algebraicly independent conformal cross ratios 
\be r_{tu}^{vs}:=\frac{(x_v-x_s)}{(x_v-x_t)}\frac{(x_t-x_u)}{(x_s-x_u)}\,.
\ee
The exponents $c_{ij}$ must fulfill the consistency conditions
\be
\sum_{j=1\atop j\neq i}^N c_{ij}=2n_i\,,\ c_{ij}=c_{ji}\,,\ 1\leq i\leq N\,.
\ee
These conditions do not fully determine the exponents $c_{ij}$ in the
case of $N\!\geq\!4$\,.
Hence, in conventional conformal field theory four-point functions and
higher 
N-point functions are not fully determined by conformal covariance.
 
In the case of conformal two-point functions and conformal three-point functions, our
strategy to construct pointlike localized correlation functions was
the following: First, we proved that the algebraic
correlation functions decrease in position space as fast as
 the associated correlation functions in conventional field theory,
which are uniquely determined by conformal covariance.
 Then, we
 transferred this property by Fourier transformation into regularity
 properties in momentum space. Finally, we were able to prove that
 the
  limit $\lambda\!\downarrow\!0$ of canonically scaled algebraic
  correlation functions converges to (a multiple of) the associated
 pointlike
 localized correlation functions in conventional conformal field theory.

 In the case of four-point functions and higher N-point functions, the situation has changed and 
we cannot expect to be able to fully determine  the form of
the pointlike localized limit in this construction, since for $N>4$
the correlation functions in conventional 
 conventional field theory are not any longer uniquely determined by conformal
 covariance. 

Beginning with the discussion of the general case with $N\!\geq\!4$\,,
we consider
algebraic N-point functions 
\be
(\,\OM,\,\left(\prod_{1\leq i\leq
  N}\,U(-x_i)\,A_i\,U(x_i)\right)\,\OM\,)
\ee
of local observables $A_i$ with minimal conformal dimensions $n_i$\,,
$i=1,2,...\,,N$\,, in a chiral theory with conformal covariance. We
want to examine the pointlike limit of canonically scaled correlation functions
\be
\lim_{\lambda\downarrow 0}\l^{-\left(\sum_{1\leq i\leq N}n_i\right)}\
(\,\OM,\,\left(\prod_{1\leq i\leq N}\,U(-
\frac{x_i}{\lambda})\,A_i\,U(
\frac{x_i}{\lambda})\right)\,\OM\,)\,.
\label{i}
\ee

Our procedure in the construction of pointlike localized
N-point functions for $N\!\geq\!4$ will be the following: We consider all possibilities to
form a set of exponents $c_{ij}$ fulfilling the
consistency conditions 
\be
\sum_{j=1\atop j\neq i}^N c_{ij}=2n_i\,,\ c_{ij}=c_{ji}\,,\ i=1,2,3,...\,,N\,.
\ee
For each consistent set of exponents a bound on algebraic N-point functions in
position space can be proved. Each single bound on algebraic
N-point functions in position space can be transferred
into a regularity property of algebraic N-point functions in momentum
space. 
We can use the same techniques as in the case of three-point functions.
Finally, we will
control the canonical scaling limit in (\ref{i}) and construct pointlike localized
 conformal N-point functions.

We present the following generalization of the conformal
cluster theorem proved above (cf.\ \cite{FrJ}) to algebraic 
N-point functions of local observables:  

\medskip

{\bf Theorem:} Let $(\A(I))_{I\in\KKK}$ be a conformally covariant local net on $\R$\,.
Let       $a_i,b_i\in\R\,,\ i=1,2,3,...\,,N$\,, and
              $a_i<b_i<a_{i+1}<b_{i+1}$ for $i=1,2,3,...\,,N\!-\!1$\,.
Let $A_i\in\A(\,(a_i,b_i)\,)$\,, $n_i\in\N$\,, and 
\be
P_k\,A_i\,\OM=P_k\,A_i^*\,\OM=0\,,\;k<n_i\,,\;i=1,2,3,...\,,N\,.
\ee
 $P_k$ here denotes the
projection on the subrepresentation of $U(SL(2,R))$ with conformal dimension $k$\,.
We then have for each set of exponents $c_{ij}$ fulfilling
the consistency conditions
\be
\sum_{j=1\atop j\neq i}^N c_{ij}=2n_i\,,\ c_{ij}=c_{ji}\,,\ i=1,2,3,...\,,N\,,
\ee
 the following bound:
\beam
\lefteqn{
|(\,\OM,\,\left(\prod_{1\leq i\leq N}A_i\right)\OM\,)|}\nn\\
&\leq&
\left(\prod_{1\leq i<j\leq N}\left|\frac{(a_i-b_i)+(a_j-b_j)}{(a_j-a_i)+(b_j-b_i)}\right|^{c_{ij}}\right)
\;\prod_{1\leq i\leq N}\|A_i\|\,.
\eeam
If we additionally assume
\be
a_1-b_1=a_2-b_2=...=a_N-b_N\,,
\ee
 we can introduce
conformal cross ratios and get
\beam
\lefteqn{
|(\,\OM,\,\left(\prod_{1\leq i\leq N}A_i\right)\OM\,)|}\nn\\
&\leq&
\left(\prod_{1\leq i<j\leq N}\left(\frac{(a_i-b_i)\,(a_j-b_j)}{(a_i-a_j)\,(b_i-b_j)}\right)^{c_{ij}/2}\right)
\;\prod_{1\leq i\leq N}\|A_i\|\,.
\eeam

\medskip

{\bf Proof:} If we pay attention to the obvious modifications needed for the
additional variables, we can use in this proof the assumptions,
 the notation, and the line of argument introduced in the
proof of the cluster theorem in the case of three-point functions.  \\
We choose an arbitrary set of exponents $c_{ij}$ fulfilling
the consistency conditions
\be
\sum_{j=1\atop j\neq i}^N c_{ij}=2n_i\,,\ c_{ij}=c_{ji}\,,\ i=1,2,3,...\,,N\,.
\ee
Let $R>0$\,. We consider the generator ${\rm \bf H}_R$ of the
following one-parameter
subgroup of $SL(2,\R)$\,:
\be
g_t\,:\,x\longmapsto\frac{x\,
                       \mb{cos}\frac{t}{2}+R\,\mb{sin}\frac{t}{2}}
{-\frac{x}{R}\,\mb{sin}\frac{t}{2}+\mb{cos}\frac{t}{2}}\,.
\ee
We know that ${\rm \bf H}_R$ is within each subrepresentation
of $U(SL(2,R))$
                                 unitarily equivalent to the
conformal Hamiltonian ${\rm \bf H}$. Therefore, the spectrum of $A_i\,\OM$
and $A^*_i\,\OM$ with respect to ${\rm \bf H}_R$ is bounded from below by $n_i$\,, $i=1,2,...\,,N$\,.
Let $0<t^-_{ij}<t^+_{ij}<2\pi$ such that 
\be
g_{t^-_{ij}}(b_i)=a_j
\ee
 and 
\be
g_{t^+_{ij}}(a_i)=b_j\,,
\ee
for  $i,j=1,2,...\,,N$\,, $i<j$\,.
We introduce
\be
F(z_1,...\,,z_N)\ :=\ (\,\OM,\,\left(\prod_{i=1}^N z_{p(i)}^{-{\rm \bf
    H}_R}\,A_{p(i)}\,
z_{p(i)}^{{\rm \bf H}_R}\right)\,
\OM\,)
\ee
in a domain of definition given by 
\be
|z_{p(1)}|<|z_{p(2)}|<...<|z_{p(N)}|
\ee
 with permutations 
$(\,p(1),p(2),...\,,p(N)\,)$ of $(\,1,2,...\,,N\,)$\,. 
This definition can uniquely be extended 
in analogy to the case of three-point functions to boundary points 
with $|z_{j}|=|z_{k}|$\,, $j,k=1,2,...\,,N$\,, $j\neq k$\,,
if 
\be
g_{t_k}([a_k,b_k])\cap g_{t_j}([a_j,b_j])\neq\emptyset\,,
\ee
thereby introducing 
\be
t_i:=-i\log z_i\,,\ i=1,2,...\,,N\,.
\ee
The line of argument presented above in the case of three-point functions
and developed for general Wightman functions in \cite{StW,Jos} proves that
 this continuation is
still an analytic function. We then define 
\be
G(z_1,...\,,z_N)\ :=\ F(z_1,...\,,z_N)\,\prod_{1\leq i<j\leq N}\,(\frac{z_i}{z_j}-z^0_{ij})^{c_{ij}/2}\,(\frac{z_j}{z_i}-z^0_{ji})^{c_{ji}/2}\,,
\ee
using the abbreviation 
\be
z_{ij}^0:=e^{i(t^-_{ij}+t^+_{ij})/2}\,,\ i,j=1,2,...\,,N\,.
\ee
This function is constructed such that with the consistency conditions
for $c_{ij}$ and with the bound on the spectrum of
${\rm \bf H}_R$ we get the following result in analogy to the cluster theorem for
three-point functions:
At the boundary points $z_i=0$ and $z_i=\infty$\,, $i=1,2,...\,,N$\,, the function
                   $G$ is bounded 
 and
can therefore be analytically continued.
 As in the case of three-point functions, we get with the maximum principle for
analytic functions further estimates
on $G$\,:  Iterating the
well-known  maximum
principle argument for the single variables, one obtains
\be
|G(1,...\,,1)|\ \leq\  \mbox{sup}_B\
|G(z_1,...\,,z_N)|\,,
\ee
where $B$ denotes the set of boundary points
\be
B\ :=\ 
\{\,|z_{j}|=|z_{k}|\ |\ g_{t_k}([a_k,b_k])\,\cap\,
  g_{t_j}([a_j,b_j])
\neq\emptyset\,,\ j\neq k
\}
\ee
with
 $t_i=-i\log z_i$\,, $i=1,2,...\,,N$\,.
The supremum of the boundary values of
$G$ can be calculated in full analogy to the 
case of the three-point functions and to the proof of the conformal
 cluster theorem (cf.\ \cite{FrJ}).  We obtain straightforward:
\beam
|(\,\OM,\,\left(\prod_{1\leq i\leq N}A_i\right)\OM\,)|
&\leq&
\left(\prod_{1\leq i\leq N}\|A_i\|\right)\,
\prod_{1\leq i<j\leq N}\,\left|\frac{\mb{sin}\frac{t^-_{ij}-t^+_{ij}}{4}}{\mb{sin}
\frac{t^-_{ij}+t^+_{ij}}{4}}\right|^{c_{ij}}.\;\;
\eeam
This estimate
converges in the limit $R\downarrow 0$ with 
\be
\lim_{R\rightarrow\infty}R\,t^-_{ij}=2(a_j-b_i)
\ee
and
\be
\lim_{R\rightarrow\infty}R\,t^+_{ij}=2(b_j-a_i)
\ee
for  $i,j=1,2,...\,,N$
to the first bound asserted 
in the theorem. If we assume 
\be
a_1-b_1=a_2-b_2=...=a_N-b_N\,,
\ee
 we
find 
\be
\left(\frac{t^-_{ij}-t^+_{ij}}{t^-_{ij}+t^+_{ij}}
\right)^2=\frac{(a_i-b_i)\,(a_j-b_j)}{(a_i-a_j)\,(b_i-b_j)}=r_{ij}\,,\;\;\;\,
i,j=1,2,...\,,N\,,
\ee
and
get the second bound. Hence, the theorem is proven.\hfill $\Box$

\medskip

For each consistent set of exponents
$c_{ij}$\,, $i,j=1,2,3,...\,,N$\,, we have proved a different bound on conformal
 four-point functions of chiral local observables. Hence, we know that
the algebraic N-point function 
\be
(\,\OM,\,\left(\prod_{1\leq i\leq
  N}\,U(-x_i)\,A_i\,U(x_i)\right)\,\OM\,)
\ee
decreases in position space at
least as fast
as the set of associated pointlike N-point functions known from conventional conformal field theory. Therefore,
the Fourier transform 
of the algebraic
N-point function has to be
 at least as regular in momentum space
as the Fourier transforms  of the associated pointlike
N-point functions
known from
conventional conformal field theory.

Technically, we follow the line of argument in the case of
three-point functions and use the formula
\be
{\cal F}(\mbox{Pol}(X)S)=\mbox{Pol}\left(\frac{\partial}{\partial Y}\right){\cal F}S
\ee
for arbitrary temperate distributions $S$ and polynomials Pol$(\cdot)$
with a (multi-dimensional)
variable $X$ in position space and an appropriate associated differential operator
$\frac{\partial}{\partial Y}$ in momentum space.
 $\F$ denotes the Fourier transformation from position space to
 momentum space.

Now, we choose $S$ to be an algebraic N-point function 
\be
(\,\OM,\,\left(\prod_{1\leq i\leq
  N}\,U(-x_i)\,A_i\,U(x_i)\right)\,\OM\,)
\ee
of local observables $A_i$ with minimal conformal dimensions
$n_i\,,\ i=1,2,...\,,N\,,$
and $X$ to be a tuple of $N-1$ algebraicly independent difference
variables
out of $x_i-x_j$\,, $i,j=1,2,...\,,N\,.$ 
The estimates in the 
cluster theorem proved above imply, that appropriate homogeneous polynomials $\mbox{Pol}(X)$ of degree
\be
\mbox{deg}(\mbox{Pol})\ =\ \left(\sum_{i=1}^N n_i\right)
-2N+2
\ee can be found such that the product $\mbox{Pol}(X)\,S$ is
still absolutely integrable in position space. 
We then see that 
$\mbox{Pol}(\frac{\partial}{\partial Y}){\cal F}S$ is continuous and bounded in
momentum space.
By locality and the condition of positive energy, the Fourier transform $F$
of an arbitrary (truncated) algebraic N-point function is known to be of the form
\be
F(p_1,...\,,p_{N-1})\ =\ G(p_1,...\,,p_{N-1})\ \prod_{i=1}^{N-1}\,\Theta(p_i)\,,
\ee
where $G$ denotes a sum of restrictions
of appropriate analytic functions to subsets of momentum space (cf.\ 
the case of three-point functions in the section above). 
One can now proceed in analogy to the argumentation in the case of
three-point functions: In a situation with conformal covariance
and minimal conformal dimensions $n_i$\,, $i=1,2,...\,,N$\,, 
the function  $G$
 can be expressed as the product of an appropriate
 homogeneous polynomial $P$ of degree 
\be
\mbox{deg}(P)\ =\ \left(\sum_{i=1}^N n_i\right)
-N+1
\ee
 and an appropriate function $H$\,, where $H$ denotes another sum of
restrictions of analytic functions to subsets of momentum space.   
Hence, we have proved that the Fourier transform of the algebraic
N-point function
\be
(\,\OM,\,\left(\prod_{1\leq i\leq
  N}\,U(-x_i)\,A_i\,U(x_i)\right)\,\OM\,)\,
\ee
can be written as 
\be
F(p_1,...\,,p_{N-1})\ =\ P(p_1,...\,,p_{N-1})\ H(p_1,...\,,p_{N-1})\ 
\prod_{i=1}^{N-1}\,\Theta(p_i)
\ee
with an appropriate homogeneous function $P$ of degree 
\be
\mbox{deg}(P)\ =\ \left(\sum_{i=1}^N n_i\right)
-N+1
\ee
 and
an appropriate continuous and bounded function $H$\,.

Using this result, we can now show in full analogy to the procedure in the last
 section
 that by
canonically scaling an algebraic N-point function we construct a
sequence of distributions that converges to an appropriate pointlike
localized N-point function of conventional conformal field theory:
\beam
&&\nn\\
\lefteqn{ \lim_{\lambda\downarrow 0}\l^{-\left(\sum_{1\leq i\leq
      N}n_i\right)}
\
(\,\OM,\,\left(\prod_{1\leq i\leq N}\,U(-
\frac{x_i}{\lambda})\,A_i\,U(
\frac{x_i}{\lambda})\right)\,\OM\,)}  
\nn\\
&&\nn\\
&=&\lim_{\lambda\downarrow 0}\l^{-\left(\sum_{1\leq i\leq
      N}n_i\right)}
\ {\cal F}_{p_i\rightarrow x_i-x_i+1}\
F(\l p_1,...\,,\l p_{N-1})\,\l^{N-1}\,\prod_{1\leq i\leq N-1}dp_i\nn\\
&&\nn\\
&=&\lim_{\lambda\downarrow 0}
\ {\cal F}_{p_i\rightarrow x_i-x_i+1}\
P(p_1,...\,,p_{N-1})\ 
H(\l p_1,...\,,\l p_{N-1})\
\prod_{1\leq i\leq N-1}\ \Theta(p_i)\ dp_i\nn\\
&&\nn\\
&=&\left(\prod_{1\leq i<j\leq N}\frac{1}{(x_j-x_i+i\e)^{c_{ij}}}\right)\;  
f(r_{t_1u_1}^{v_1s_1},...\,,r_{t_{N-3}u_{N-3}}^{v_{N-3}s_{N-3}})\,.\\
&&\nn
\eeam
Again,  $f(\cdot,...\,,\cdot)$
denotes an appropriate function
depending on $N\!-\!3$ algebraicly independent conformal cross ratios 
\be r_{tu}^{vs}:=\frac{(x_v-x_s)}{(x_v-x_t)}\frac{(x_t-x_u)}{(x_s-x_u)}\,.
\ee
The exponents $c_{ij}$ must fulfill the consistency conditions
\be
\sum_{j=1\atop j\neq i}^N c_{ij}=2n_i\,,\ c_{ij}=c_{ji}\,,\ 1\leq i\leq N\,,
\ee
 which
do not fully determine the exponents.
Hence, the general form of the pointlike localized conformal correlation
functions constructed from algebraic quantum field theory has been
determined to be exactly the general form of the N-point functions known from
conventional conformal field theory. In both approaches conformal
covariance
does not fully determine the form of N-point functions for $N>4$\,.
\subsection{Wightman Axioms and Reconstruction Theorem}
The most common axiomatic system for pointlike localized quantum
fields is the formulation of Wightman axioms given in \cite{StW} and
\cite{Jos}. (If braid group
statistics has to be considered and the Bose-Fermi alternative does not hold in general, the classical formulation of
\cite{StW} and \cite{Jos} has to be modified for the charged case by introducing the axiom of weak locality
instead of locality \cite{FRS1,FRS2}.)

 The
construction of pointlike localized correlation functions in this paper uses sequences of algebraic correlation functions of local 
observables.
The algebraic correlation functions obviously fulfill positive
definiteness, 
conformal covariance, locality, and the spectrum
condition. 
 Hence,
if the sequences converge, the set of pointlike limits of 
algebraic correlation functions fulfills the Wightman axioms (see
\cite{StW}) by
construction. 
By the reconstruction theorem in \cite{StW} and \cite{Jos}, the
existence of Wightman fields associated with the Wightman functions is
 guaranteed and this Wightman field theory is unique up to unitary
 equivalence. 

We do not know at the moment whether the Wightman fields
 can be identified with the pointlike localized field operators
 constructed in \cite{FrJ} from the
 Haag-Kastler theory.  We do
 not know either whether the Wightman fields are affiliated to the
 associated von Neumann algebras of local observables and how the Haag-Kastler net we
 have been starting from can be reconstructed from the Wightman
 fields. Possibly, the Wightman fields cannot even be realized in the
 same Hilbert space as the Haag-Kastler net of local observables.

We do know, however,  that the Wightman theory associated with
the Haag-Kastler theory is non-trivial: The two-point functions of this
Wightman fields are, by construction,
identical with the two-point functions of the pointlike localized
field operators constructed in \cite{FrJ}.
And
we have already proved that those pointlike field vectors can be chosen to
be non-vanishing and that the vacuum vector is cyclic for a set of all 
field operators localized in an arbitrary interval.

 It shall be
pointed out again that those pointlike fields constructed in
\cite{FrJ,Joe3} could not be proved to fulfill the Wightman axioms, since we were not
able to find a domain of definition that is stable under the
action of the field operators.

To summarize this paper, we state that starting from a chiral conformal
Haag-Kastler theory we have found a canonical construction of
 non-trivial Wightman fields. The reconstruction of
 the original net of von Neumann algebras of local observables from the Wightman fields
 could not explicitly be presented, since we do not know whether the
 Wightman fields can be realized in the same Hilbert space as the
 Haag-Kastler net. 

Actually, Borchers and Yngvason \cite{BoY} have investigated similar
 situations and have shown that such problems can occur in quantum field theory. In \cite{BoY} the question is discussed under which
 conditions a Haag-Kastler net can be associated with a Wightman
 theory. 
The condition for the locality of the associated algebra net turned out be
a property of the Wightman fields called ``central positivity". Central positivity is
fulfilled for Haag-Kastler nets and is stable under pointlike limits \cite{BoY}.
Hence, the Wightman fields constructed in this thesis fulfill 
central positivity.  
The
 possibility, however, that the local net has to be defined in an enlarged
 Hilbert space could not be ruled out in general by
 \cite{BoY}.  

Furthermore, it has been proved in \cite{BoY} that Wightman fields
fulfilling generalized H-bounds (cf.\ \cite{DSW}) have associated
local nets of von Neumann algebras
that can be defined in the same Hilbert space. The closures of the
Wightman field operators are then affiliated to the associated local
algebras.
We could not prove generalized H-bounds for the Wightman fields
constructed in this thesis. Actually, we suppose that the
criterion of generalized H-bounds is too strict for general conformal --
and therefore massless -- quantum field theories. (Generalized) H-bounds have been proved,
however,
for massive theories, i.e.\ for models in quantum field theory with
massive
 particles (cf.\ also \cite{DrF,FrH,Sum,Buc1}).
\paragraph{Acknowledgements}\nichts\\
This paper is one part of the author's dissertation. 
We would like to thank Prof.\,Dr.\,K.\,Fre\-den\-ha\-gen for his
confidence, constant encouragement, and the numerous inspiring
discussions over the whole period of the work. I am indebted to him
for many important insights I received by his guidance. His
cooperation was crucial and fruitful for this work. 

The financial support given by the Friedrich-Ebert-Stiftung is
gratefully acknowledged.
\end{document}